\documentclass[%
 reprint,
 amsmath,amssymb,
 aps,
 pre,
 superscriptaddress,
]{revtex4-2}

\usepackage{graphicx}
\usepackage{dcolumn}
\usepackage{bm}
\usepackage{enumitem}

\begin{document}

\preprint{APS/123-QED}

\title{Stochastic Resonance in a Thermally Driven Low-Dimensional Geodynamo Model}

\author{Giuseppina Nigro}
\email{giuseppina.nigro@roma2.infn.it}
\affiliation{%
Dipartimento di Fisica, Universit\`a degli Studi di Roma Tor Vergata, Roma, Italy
}%
\author{Edoardo Cascio}%
\altaffiliation{Current address: Department of Environmental Sciences, Informatics and Statistics, Ca' Foscari University of Venice, Venezia, Italy; also at INAF-Istituto di Astrofisica e Planetologia Spaziali, Roma, Italy}
\affiliation{%
Dipartimento di Fisica, Universit\`a degli Studi di Roma Tor Vergata, Roma, Italy
}%

\author{Giuseppe Consolini}
\affiliation{
INAF-Istituto di Astrofisica e Planetologia Spaziali, Roma, Italy
}%

\author{Francesco Berrilli}
\affiliation{%
 Dipartimento di Fisica, Universit\`a degli Studi di Roma Tor Vergata, Roma, Italy
}%

\date{\today}

\begin{abstract}

Geomagnetic field reversal sequences exhibit persistence times spanning a broad range, from a few $10^4$ years to superchrons lasting more than $10^7$ years. 
Despite extensive observational and theoretical work, the physical mechanisms governing how such reversals occur and how their broad temporal variability is organized are still not fully understood.

Here we investigate the temporal variability of geomagnetic polarity in a thermally driven low-dimensional geodynamo model subject to a slow periodic modulation of the control parameter governing the large-scale induction, namely the $\alpha$-effect parameter. We find that the modulation generates a multipeaked probability density function of magnetic persistence times, with local maxima occurring at approximately integer multiples of the modulation timescale, as expected in a stochastic-resonance-like regime. The peak positions follow an approximately linear dependence on their index, showing that the characteristic timescales selected by the system are set by the imposed modulation period. These results provide a physically motivated numerical framework in which slow modulation of a geodynamo control parameter can organize reversal statistics through stochastic-resonance-like dynamics.

\end{abstract}

\maketitle


\section{Introduction}

The Earth has a magnetic field that, on the ground and outside of the planet, is largely dipolar, with the dipole axis aligned with the rotation axis to within $\lesssim 10^{\circ}$. It is generated by the geodynamo, driven by convection in the electrically conducting liquid outer core, which is sustained by buoyancy sources associated with secular cooling, inner-core growth, and heat flow across the core–mantle boundary (CMB).
Paleomagnetic records show that the Earth's magnetic field exhibits variations over different timescales, from months to billions of years. 
One of the most remarkable features of geomagnetic field variability is the occurrence of polarity reversals. During a reversal, the strength of the dipole component diminishes, and the dipole axis may swing rapidly a few times. The dipole then reverses its direction, and within about $10 \; \mathrm{kyr}$ the entire process is completed and the magnetic field reversed. 
The process may be interrupted, causing the dipole component to revert to its original polarity, resulting in what is called a magnetic excursion.

The typical duration of geomagnetic reversals is a few thousand years, which is much shorter than the typical time interval between successive reversals ranging from $10^4$ to $10^7$ years \cite{Cox_JGR_1968, CK95, Valet_Nature_2005, Valet_Fournier_RvGeo_2016}.
This strong separation of time scales indicates that the geodynamo spends most of its time in relatively stable polarity states, occasionally interrupted by rapid transitions and/or polarity excursions.
The long-lived states, during which the dipolar component of the Earth's magnetic field maintains the same polarity, are associated with the residence times (or persistence times) of the system in each polarity, whereas reversals correspond to transitions between them.

Long-term geomagnetic variations have been the subject of a wide literature, exhibiting near-periodicity that has been interpreted as the coupling between Earth internal dynamics and orbital parameters \citep{Channell_et_al_Nature_1998, Yamazaki_Oda_Science_2002, Yamazaki_PEPI_2002, Nakagawa_Tackley_GGG_2005}. 

Consolini \& De Michelis \citep{consolini2003stochastic}, analyzing persistence time statistics from the merging of two scales compiled by Cande and Kent \citep{cande1995revised} and Ogg \citep[in][]{ahrens1995global}, proposed stochastic resonance (SR) as the mechanism for the occurrence of geomagnetic polarity reversals \citep{benzi1981mechanism, gammaitoni1998stochastic}. 

SR, originally proposed as a possible mechanism of the quasi-periodicity of ice ages, is a special class of noise-induced cooperative phenomenon. There are three basic elements of SR: {\it i) }a double well potential or a bistable system, {\it ii)} a weak coherent input modulating the double well or the probability transition between the two states, and {\it iii)} a noise source. The name stochastic resonance derives from the fact that an optimal level of noise amplitude can enhance the synchronous transitions with coherent modulation of the potential minima. The optimal amount of noise for the synchronization is observed when the average waiting time, $T_k$, between the interwell transition between the two-states is half the period $T_\omega$ of coherent modulation, i.e., $2T_k \simeq T_\omega$. In this sense, SR is an example of a cooperative noise-induced transition process. 

Consolini \& De Michelis \citep{consolini2003stochastic} have shown that the multipeaked character of the distribution function of the polarity time intervals can be explained in terms of a SR phenomenon, characterized by an average waiting time for reversals of the order of $\sim 100$ kyr, a periodicity that suggests a possible link between geomagnetic polarity reversals and the variation of the Earth’s orbital eccentricity.
Subsequently, Lorito et al. \citep{Lorito_et_al_2005AN} showed that SR might be a plausible mechanism to explain geomagnetic polarity reversals by applying periodic modulations to an axisymmetric mean-field $\alpha\Omega$-dynamo model. 

Although an $\alpha\Omega$-dynamo has often been considered a plausible scenario \citep{Ryan_Sarson_2011PEPI}, several numerical and theoretical studies suggest that the $\alpha^2$ or $\alpha^2\Omega$ dynamo mechanisms may be a more appropriate framework to account for the stability of the geomagnetic dipole and the occurrence of geomagnetic polarity reversals. The use of an $\alpha^2$ dynamo framework for the geodynamo is physically motivated by the rapidly rotating helical nature of the outer-core convection. In Earth-like dynamo regimes, convection tends to organize into columnar structures aligned with the rotation axis, as supported by geodynamo simulations  
\citep{Glatzmaier_Roberts_PEPI_1995, Aubert_et_al_2017, Lin_et_al_Nature_2025}. The flows with mirror-symmetry breaking associated with these columnar structures can generate both poloidal and toroidal large-scale magnetic field components through the $\alpha$-effect alone \citep{Steenbeck_Krause_1969, Moffatt_1978}.

Numerical geodynamo simulations have shown that, in strongly rotationally constrained regimes characterized by columnar convection, the conversion between magnetic-field components can occur predominantly through the $\alpha$-effect in both directions, corresponding to an $\alpha^2$ dynamo \citep{Busse_JFM_1970, Christensen_et_al_1999, Olson_Christensen_Glatzmaier_JGR_1999, Schrinner_et_al_2007, Yadav_PNAS_2016, Schaeffer_GeoJI_2017}, whereas different convective regimes coupled with rotation may involve additional contributions from differential rotation, corresponding to an $\alpha^2\Omega$ dynamo \citep{Kono_Roberts_2002, Yadav_et_al_GeoJI_2016}, when the differential rotation remains coherent in space and time \citep{Nigro_et_al_2017}.
Mean-field dynamo models further support the idea that $\alpha^2$-type descriptions capture essential features of dipole generation and polarity variability in convection-driven dynamos \citep{Jones_Longbottom_Hollerbach_1995, Rudiger_et_al_2003, Giesecke_et_al_2005, Stefani_Gerbeth_PhRvL_2005, Stefani_et_al_E&PSL_2006, Sorriso_et_al_2007PEPI}. 

Therefore, since the above-mentioned theoretical and numerical evidence does not support the presence in the Earth's outer core of a significant velocity shear, coherent in time and space, the role of an $\Omega$ mechanism can be neglected in the geodynamo formulation.
For this reason, a low-dimensional $\alpha^2$-dynamo framework can provide a useful and physically grounded approximation to  investigate bistability, metastability, and reversal dynamics in geodynamo.

Within this framework, the large-scale geomagnetic field can be interpreted as the outcome of a nonlinear dynamical system characterized by multiple metastable states, in which turbulent fluctuations and slowly varying external conditions may induce transitions between polarity states. This perspective naturally motivates the adoption of reduced dynamical models, where bistability is explicitly represented, and the interplay between the modulation of the bistability structure and turbulent fluctuations can give rise to reversal sequences through mechanisms akin to stochastic resonance.

Motivated by the above considerations, in this work we investigate whether a statistical organization similar to that observed by Consolini \& De Michelis for geomagnetic reversals \citep{consolini2003stochastic} can arise in a thermally driven low-dimensional $\alpha^2$-dynamo model. In particular, we introduce a slow time modulation of the parameter controlling the dynamo efficiency and analyze how this modulation affects the persistence times of the large-scale magnetic polarity state. However, in the present work the slow modulation is not intended as a detailed representation of orbital forcing. Rather, it is introduced as a controlled way to investigate whether slow variability in the $\alpha$ effect can organize the statistics of polarity transitions in a stochastic-resonance-like manner.

Because of the limited knowledge of some fundamental parameters controlling Earth's outer-core dynamics and because direct numerical simulations cannot yet fully access the extreme parameter regime of the geodynamo, the purpose of the present work is to reproduce geomagnetic polarity reversals in terms of their dynamical behavior, temporal patterns, and statistical properties. 
The exploratory investigation presented here is therefore aimed at identifying possible mechanisms that may play a key role in the dynamics of geomagnetic reversals. For this purpose, a low-dimensional model is particularly useful, since it can generate sufficiently rich reversal statistics while retaining the simplicity required to isolate specific ingredients that enter into play in the dynamics under investigation.

\section{Model Description}

Our approach adopts a thermally driven magnetoconvective shell model, modified at the largest scale (i.e., the smallest shell) by introducing a term for the $\alpha$ effect that also includes the $\alpha$-quenching (see \citep{Nigro_Carbone_2011, Nigro_2013GApFD, Nigro_2022, Nigro_2025}, and for the classical GOY shell model \citep[][]{Gledzer_1973, Ohkitani_Yamada_1989}). 

The use of a shell model is motivated by the need to retain nonlinear turbulent interactions over a broad range of scales, while keeping the system sufficiently simple to allow long integrations and statistical analysis of reversal sequences. This approach necessarily neglects several aspects of the complete geodynamo problem, including spherical geometry, realistic boundary conditions, and the detailed spatial structure of core convection. However, it preserves the ingredients that are essential for the mechanism investigated here: turbulent fluctuations, magnetic induction, thermal driving, and a large-scale magnetic degree of freedom with two preferred polarity states.

Therefore, the model equations are written as: 

%
\begin{widetext}
\begin{eqnarray}
\hspace*{-0.7cm} \left(\frac{d}{dt} + \nu k_n^2\right) u_n &=&  - \tilde{\alpha} \theta_n + {i} k_n[(u_{n+1}u_{n+2}-b_{n+1}b_{n+2}) \!
- \! \frac{\epsilon}{2}(u_{n-1}u_{n+1}-b_{n-1}b_{n+1}) \!
- \! \frac{1-\epsilon}{4}(u_{n-2}u_{n-1}-b_{n-2}b_{n-1})]^* 
\label{eq:GOY_u}
\\
\hspace*{-0.7cm} \left( \frac{d}{dt} +  \chi k_n^2\right) \theta_n &=&
{i} k_n [\alpha_1 u_{n+1}^*\theta_{n+2}^* + \alpha_2 u_{n+2}^* \theta_{n+1}^* 
+ \beta_1 u_{n-1} \theta_{n+1} - \beta_2 u_{n+1} \theta_{n-1} 
+ \gamma_1 u_{n-1} \theta_{n-2} + \gamma_2 u_{n-2} \theta_{n-1}]^* \!+ \! f_n
\label{eq:Thermal}
\\
\hspace*{-0.7cm} \forall &n& \in \{1, 2, \dots, N \}
\nonumber
\\
\hspace*{-0.7cm} \frac{d b_1}{dt} + \eta k_1^2 b_1 = 
&{ i}& \frac{k_1}{6} \; (u_{2}^*b_{3}-b_{2}^*u_{3})+ \mu(t) b_1 \left( 1- \frac{b_1^2}{B_0^2}\right) 
\label{eq:b1}
\\
\hspace*{-0.7cm} \left(\frac{d}{dt} + \eta k_n^2\right) b_n  &=&
{i} k_n [(\!1\!\!-\!\epsilon\!-\!\epsilon_m\!)(\!u_{n+1}b_{n+2}\!-\!b_{n+1}u_{n+2})
\!+ \! \frac{\epsilon_m}{2}(\!u_{n-1}b_{n+1}\!-\!b_{n-1}u_{n+1}) \!+ \!
\frac{1-\epsilon_m}{4}\!(\!u_{n-2}b_{n-1}\!-\!b_{n-2}u_{n-1})]^*
\label{eq:GOY_b}
\\
\hspace*{-0.7cm} \forall &n& \in \{2, \dots, N \}
\nonumber
\end{eqnarray}
\end{widetext}
%
\begin{figure*}
   \includegraphics[width=18cm]{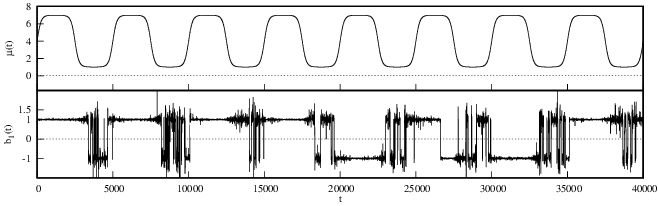}
    \caption{The large-scale geomagnetic field $b_1(t)$ (bottom panel) is more prone to invert its polarity, switching between 1 and -1 values, during the time interval of $\mu(t)$ minima. This simulation is obtained keeping constant $\chi=\eta=\nu=10^{-4}$, $\tilde{\alpha}=0.5$, and finally $\mu_c=4$, $\mu_A=3$ and $T_{\omega}=5000$ for the $\mu(t)$ function that shapes the $\alpha$-dynamo effect.}
    \label{fig:b1_Than}
\end{figure*}
%
where $u_n$ are the velocity field fluctuations, $b_n$ magnetic field fluctuations, and $\theta_n$  temperature fluctuations in the shell $n$, with $n=1, \dots, N$, where $N$ is the total number of shells, $\nu$ is the kinematic viscosity, $\eta$ the magnetic diffusivity, $\chi$ is the thermal diffusivity, $\tilde{\alpha}$ is the thermal convection coefficient, and finally $f_n$ is the forcing. The $\alpha$-dynamo effect is described in the model by the pitchfork bifurcation term, namely the last term in the RHS of equation (\ref{eq:b1}), where the parameter $\mu(t)$ is related to the kinetic helicity of flow motions \citep{Pouquet_et_al_JFM_1976}. We assume that $\mu(t)$ is a time-dependent function with a shape that we will discuss in the next subsection.

We assume that a level of equipartition is achieved in the saturation regime, where $B_{0}$ is the amplitude of the equipartition or the saturation magnetic field.
The values of coefficients $\epsilon = 1/2$ and $\epsilon_m = 1/3 $, as equations (\ref{eq:GOY_u})-(\ref{eq:GOY_b})-(\ref{eq:Thermal}) coincide with the MHD GOY shell model \citep{Frick_Sokoloff_1998} when $\tilde{\alpha} = 0 $ and $\theta_n = 0$ for every $n=1,...,N$. The coefficients in the temperature equations are $\alpha_1 = \alpha_2 = 1$, $\beta_1 = \beta_2 = 1/2$, and $\gamma_1 = \gamma_2 = - 1/4$ such that we adopt the coupling with the model made by \citet{Jensen_et_al_1992} (see also \cite{Mingshun_Shida_1997}).

The model equations (\ref{eq:GOY_u}), (\ref{eq:GOY_b}) and (\ref{eq:Thermal}) are written in a dimensionless form, where we measure the velocity in terms of the free-fall velocity $U = \sqrt{ \hat{\alpha} g L \Delta T}$ and measure the magnetic field in units of $ B_0 = U$, corresponding to the Alfvénic field strength. Temperature, length and time are measured in units of $\Delta T$, $L$ and the free-fall time $( L/ \hat{\alpha} g \Delta T)^{1/2}$, respectively.

The forcing function $f_n$ mimics a stochastic process with finite correlation time, obtained by integrating a Langevin-type equation acting on the largest modes ($f_n \neq 0$ for $n=1,2,3$). This prescription leads to a Gaussian process that ensures statistical stationarity while avoiding the unphysical properties of purely white-noise forcing. The presence of a finite correlation time (namely, in this case one free-fall time) allows the forcing to retain short-term temporal coherence, consistent with the characteristic timescales of large-scale convective motions (see \citet{Nigro_2025} for more details of this forcing).

The simulations are controlled by the following dimensionless parameters:
\begin{itemize}[label=\raisebox{0.2ex}{\tiny$\bullet$}, leftmargin=*, itemsep=0pt]
\item $\tilde{\alpha}$, the thermal expansion that controls the thermal driving of the flow velocity;
\item $\nu$, the kinematic viscosity;
\item $\eta$, the magnetic diffusivity;
\item $\chi$, the thermal diffusivity;
\item the amplitude of the forcing, which is set equal to 1 in a dimensionless unit;
\item $\mu(t)$ which is a time-dependent parameter discussed in the following subsection.
\end{itemize}
Based on these parameters, we define the following characteristic dimensionless numbers:
\begin{itemize}[label=\raisebox{0.2ex}{\tiny$\bullet$}, leftmargin=*, itemsep=0pt]
\item the hydrodynamic Reynolds number, $\mathrm{Re} = L u_0 / \nu$;
\item the magnetic Reynolds number, $\mathrm{Rm} = L u_0 / \eta$;
\item the Rayleigh number, $\mathrm{Ra} = \tilde{\alpha} \, \theta_0 \, L^3 / (\nu \chi)$.
\end{itemize}
where $\theta_0 = \sqrt{\langle \sum_{n=1}^N \theta_n^2 \rangle_t}$, $u_0 = \sqrt{\langle \sum_{n=1}^N u_n^2 \rangle_t}$, and $b_0 = \sqrt{\langle \sum_{n=1}^N b_n^2 \rangle_t}$ denote the rms amplitudes of temperature, velocity, and magnetic field, respectively. These quantities remain of order unity in dimensionless units for all simulations made (i.e., $\theta_0 \approx u_0 \approx b_0 \approx 1$).

\subsection{Parameter $\mu$}

Although superchrons are not strictly periodic and their number is limited, they suggest the existence of long timescales over which the geodynamo may remain in a quasi-stable polarity state. Motivated by this separation of timescales, we model the parameter $\mu(t)$ as a slowly varying periodic function, intended as an idealized representation of long-term modulations of dynamo efficiency. This choice is consistent with scenarios in which kinetic helicity may be modulated by slowly varying external forcings or boundary conditions, such as long-term variations in CMB heat flux, which, in turn, can probably be modified by Earth's orbital conditions.

In addition, the persistence of the geomagnetic field in a nearly stable state during superchrons motivates the use of a smooth saturating modulation profile, such as a hyperbolic tangent, which naturally captures extended phases of quasi-stationary behavior separated by relatively rapid transitions. 

Therefore, we assume a periodic modulation of the $\mu$ parameter according to the following equation:
\begin{equation}
     \mu(t) = \mu_c + \mu_A \; \tanh\!\left[ 3 \sin\!\left( {2 \pi t}/{T_{\omega}} \right) \right]
    \label{eq_mu_time}
\end{equation}
where $\mu_c$ is a constant of value that we set in the test cases discussed in this paper $\mu_c= 4$, and the second term is a modulation function with amplitude $\mu_A$ (which we set equal to $\mu_A = 3$). Indeed, $\mu$ has to assume values on the order of kinetic energy, which are of the order of one in dimensionless units.

The inclusion of a time-dependent modulation of the parameter $\mu$ therefore provides a controlled dynamical framework in which the role of a slowly varying dynamo efficiency can be tested in our investigation.

\section{Results}

The model equations (\ref{eq:GOY_u}), (\ref{eq:Thermal}), (\ref{eq:b1}) and (\ref{eq:GOY_b}) are numerically solved by adopting a fourth-order Runge–Kutta numerical scheme. 
As we are interested in the dynamics of the large-scale magnetic field, we focus on the time evolution of the magnetic on the first shell $b_1(t)$, which is given by the integration of equation (\ref{eq:b1}).
In this equation, we can distinguish three effects:
\begin{itemize}[label=\raisebox{0.2ex}{\tiny$\bullet$}, leftmargin=*, itemsep=0pt]
    \item the diffusivity of the largest-scale magnetic field $b_1(t)$ (second term on the LHS);
    \item the pitchfork bifurcation term (the last terms proportional to $\mu(t)$ on RHS), where the attracted fixed points are determined by the condition $b_1 = \pm B_0$ (two stable equilibria), and $b_1=0$ is an unstable equilibrium;
    \item the non linear terms that couples the nearest two neighboring shells with $b_1$ (triadic interaction) and produces a dynamical perturbation of the largest-scale magnetic field $b_1$ when it is set down on one of the two stable equilibria above mentioned. When the perturbation is large enough, $b_1$ can escape from the fixed point where it was, and eventually reaches the other fixed point. In this transition from one to the other fixed point, $b_1$ reverts its sign, thus reproducing a large-scale field polarity reversal.
\end{itemize}

In order to characterize the bistable behavior associated with polarity changes, we analyze the behavior of the real part of $b_1(t)$, which, with an abuse of notation, we still call $b_1(t)$ (i.e. $b_1(t) \equiv \Re\!\left[b_1(t)\right]$).

The typical behavior of $b_1$ in time is provided by the bottom panel of Fig.\ref{fig:b1_Than}, which depicts the simulation output obtained with $\chi=\eta=\nu=10^{-4}$, the thermal expansion coefficient $\tilde{\alpha}=0.5$, and for the parameter $\mu(t)$ reported in the top panel of the same figure, when we select $\mu_c=4$, $\mu_A=3$ and $T_{\omega}=5000$ in dimensionless units (see equations (\ref{eq:b1}) and (\ref{eq_mu_time})).

\subsection{The double-well potential}
%
\begin{figure*}[t]
  \centering
  \includegraphics[width=0.98\textwidth]{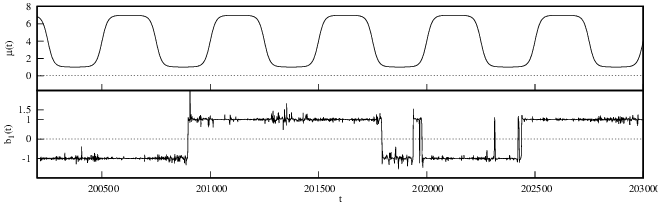}
  \caption{Time series of $b_1(t)$ (bottom panel) obtained in the simulation with parameters $\nu = \chi = \eta = 10^{-4}$, $\tilde{\alpha} = 0.5$ and the time-dependent $\mu(t)$ parameter, as depicted in top panel, with $\mu_c = 4$, $\mu_A = 3$, and $T_{\omega} = 500$.}
  \label{fig:timeseries_b1}
\end{figure*}
%
\begin{figure*}[t]
  \centering
  \includegraphics[width=0.98\textwidth]{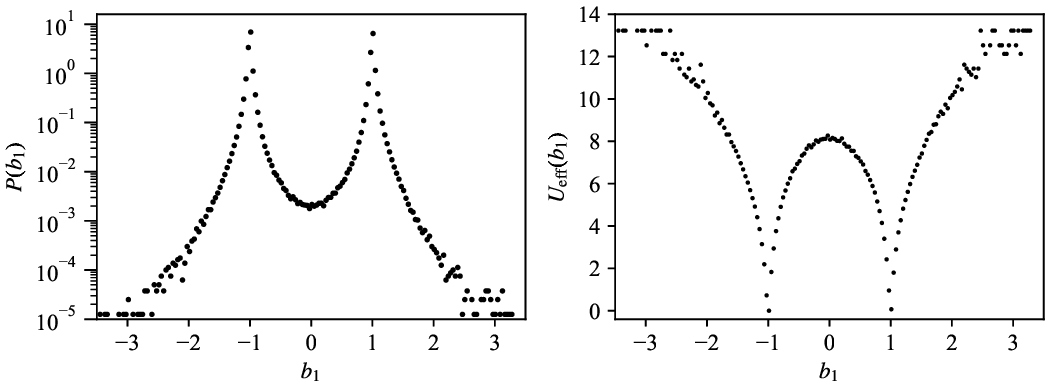}
  \caption{Probability density $P(b_1)$ (left) and the corresponding effective pseudopotential $U_{\rm eff}(b_1)=-\log P(b_1)$ (right) obtained from the simulation with parameters $\nu = \chi = \eta = 10^{-4}$ and $\tilde{\alpha} = 0.5$ for the case B with $T_{\omega} = 500$, $\mu_c = 4$ and $\mu_A = 3$.} 
  \label{fig:doublewell}
\end{figure*}

Due to the pitchfork bifurcation term, the long-term dynamics of $b_1(t)$ can be interpreted in terms of an effective double-well potential, a model already adopted in different low-dimensional descriptions of geomagnetic reversals \citep{Schmitt_et_al_2001PEPI, Lorito_et_al_2005AN}.
Indeed, at the level of an effective one-dimensional model, neglecting the fast fluctuating contribution of the triadic interaction, the reduced dynamics for $b_1$ can be written in the normal form of a supercritical pitchfork bifurcation with linear dissipation, 
\begin{equation}
\dot{b_1} = (\mu-\eta k_1^2)\,b_1 - \frac{\mu}{B_0^2}\,b_1^3,
\label{eq:x_reduced}
\end{equation}
which can be recast as an overdamped relaxation in a Landau potential $U(b_1)$ defined by $\dot{b_1}=-dU/db_1$, yielding
\begin{equation}
U(b_1)= -\frac{\mu-\eta k_1^2}{2}\,b_1^2+\frac{\mu}{4B_0^2}\,b_1^4 \,+\,{\rm const}.
\label{eq:U_landau}
\end{equation}
For $\mu>\eta k_1^2$, $U(b_1)$ exhibits a double-well structure with two stable minima at
\begin{equation}
b_1^\star=\pm B_0\sqrt{1-\frac{\eta k_1^2}{\mu}},
\end{equation}
corresponding to the two preferred polarity states, while for $\mu\le \eta k_1^2$ the potential reduces to a single well centered at $b_1=0$ (case not considered here as $\mu > \eta  k_1$ always). 
Within this reduced description, the height of the potential barrier that separates the two wells is given by
\begin{equation}
\Delta U = U(0)-U(b_1^\star)= \frac{B_0^2}{4\mu}\left(\mu-\eta k_1^2\right)^2 \underset{\eta \ll 1} \simeq {\mu} \frac{B_0^2}{4} 
\label{eq:barrier_height}
\end{equation}
demonstrating that the barrier increases and/or decreases with $\mu$.

In equation \eqref{eq:b1}, the nonlinear coupling terms appearing in the first two terms in the RHS generate fluctuations that induce transitions between the wells; accordingly, we also reconstruct a data-driven pseudopotential from the stationary statistics of $b_1(t)$ by estimating the probability density $P(b_1)$ and defining the following pseudopotential
\begin{equation}
U_{\rm eff}(b_1) = -\ln P(b_1) + {\rm const},
\label{eq:Ueff_pdf}
\end{equation}
so that the minima of $U_{\rm eff}$ identify the most probable polarity states and the barrier height quantifies the rarity of excursions across $b_1\simeq 0$.

As an illustrative example of the system's behavior in the double-well potential, we report in the left and right panels of Fig.~\ref{fig:doublewell}, respectively, the probability density $P(b_1)$ and the effective pseudopotential obtained from the simulation for the hyperbolic tangential shape of $\mu(t)$ with parameters  $\mu_c = 4$, $\mu_A = 3$ and $T_{\omega} = 500$, dissipative coefficients $\nu = \chi = \eta = 10^{-4}$ and thermal expansion coefficient $\tilde{\alpha} = 0.5$. In particular, we reconstruct the stationary probability density, i.e. PDF $P(b_1)$, of the large-scale magnetic observable $b_1(t)$ from long time series produced by our shell-model simulation (see Fig.~\ref{fig:timeseries_b1}). After discarding an initial transient (a fixed fraction of the total samples), we estimate $P(b_1)$ using a normalized histogram with bin width chosen according to the Freedman--Diaconis rule, with limits on the total number of bins to avoid over- or under-binning \citep{Freedman_Diaconis_1981}.

To characterize bistable dynamics and quantify a possible asymmetry between the two polarity states, we compute the minima, barrier heights and asymmetry of $U_{\rm{eff}}$ after fixing the additive constant in Eq.~\eqref{eq:Ueff_pdf} by shifting the minimum to zero $\min(U_{\rm eff})=0$. 

The locations of the two wells, $b_-^\star$ and $b_+^\star$, are identified as the minima of $U_{\rm eff}$ on the negative and positive sides, respectively (excluding a small neighborhood around $b_1=0$ to avoid selecting the barrier region). The location of the barrier (saddle) $b_s$ is defined as the maximum of $U_{\rm eff}$ in a narrow interval around $b_1\simeq 0$. The barrier heights with respect to each polarity state are, therefore,
\begin{equation}
\Delta U_{\pm} = U_{\rm eff}(b_s) - U_{\rm eff}(b_{\pm}^\star).
\end{equation}
A convenient measure of the polarity asymmetry of the two wells is the difference in well depths, namely
\begin{equation}
\delta U_{\rm eff} \equiv U_{\rm eff}(b_+^\star) - U_{\rm eff}(b_-^\star),
\label{eq:deltaU_fromU}
\end{equation}
which is equivalent, by construction of $U_{\rm eff}$, to the PDF-peak ratio
\begin{equation}
\delta U_{\rm pdf}
= -\ln\!\left[\frac{P(b_+^\star)}{P(b_-^\star)}\right],
\label{eq:deltaU_fromP}
\end{equation}
up to the finite-bin and regularization effects inherent to histogram-based estimates.

For the case shown in Fig.~\ref{fig:doublewell}, we find two well minima located at $b_-^\star \simeq -0.987$ and $b_+^\star \simeq 1.011$. The barrier (saddle) is located close to the origin at $b_s \simeq -0.027$. The corresponding barrier heights are $\Delta U_- \simeq 6.61$ and $\Delta U_+ \simeq 6.52$, indicating comparable activation barriers for transitions out of each polarity state. The well-depth asymmetry is small, indicating polarity bias, with $\delta U_{\rm eff} \simeq 0.09$, consistent with the PDF-peak estimate $\delta U_{\rm pdf} \simeq 0.07$.

%
\begin{figure*}[t]
    \centering
    \includegraphics[width=\textwidth]{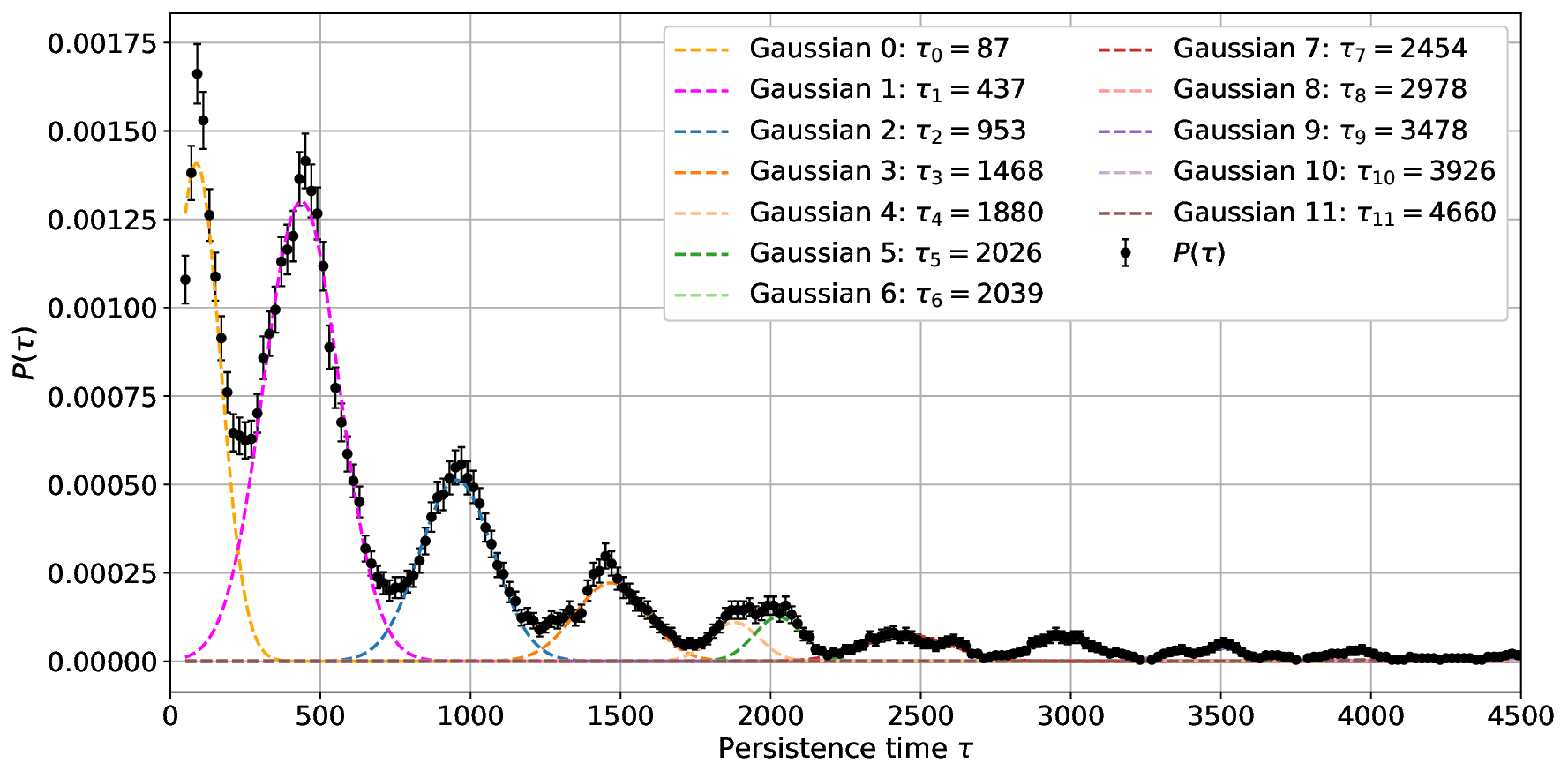}
    \caption{The features of a stochastic resonance process emerge from the probability density function $P(\tau)$ of magnetic residence times, computed using a sliding-window binning procedure (see text). Specifically, the corresponding best-fit multi-Gaussian model exhibits a series of local maxima located at approximately integer multiples of the fundamental time scale $T_{\omega}=500$, corresponding to the modulation period of the $\mu$ parameter that accounts for the action of kinetic helicity on the magnetic induction.}
    \label{fig:Ptau_gaussians}
\end{figure*}
%

We also report residence-time asymmetry, defined as the fraction of samples with $b_1(t)>0$ versus $b_1(t)<0$, as an independent diagnostic of a possible preference for polarity. Consistently, the residence-time fractions yield $f_- \equiv \langle \mathbb{I}(b_1<0)\rangle \simeq 0.526$, which means about 52.6\% of the time the system is in the left well (i.e. $b_1<0$) and $f_+ \equiv \langle \mathbb{I}(b_1>0)\rangle \simeq 0.474$, that is, approximately 47.4\% of the time the system is in the right well (i.e. $b_1>0$). Therefore, the occupation ratio is $f_+/f_- \simeq 0.90$, and thus the two-well potential is quite symmetric.

\subsection{Stochastic Resonance in the Geodynamo Model}

We analyze persistence times between geomagnetic polarity reversals $\tau$, after removing short events below a threshold to exclude the excursion \footnote{A geomagnetic excursion is a transient episode characterized by a pronounced weakening of the geomagnetic field, during which the field may temporarily assume reversed or anomalous directions, but ultimately returns to the original polarity without completing a full reversal}. We compute a binned representation $P(\tau)$ by sliding a fixed-width window $\Delta \tau$ along the $\tau$-axis and counting how many persistence times fall within each window. Specifically, the window is initially centered in $\tau_1$, such as $\tau_1- \Delta \tau = \tau_{min}$, and then moved along the $\tau$-axis with a step $\delta \tau$, so that the $i$-th window spans the interval
\begin{equation}
\left[\tau_i - \frac{\Delta\tau}{2},\ \tau_i + \frac{\Delta\tau}{2}\right)    
\end{equation}
covering the entire sample between the minimum and maximum observed persistence times, i.e. $\tau_{min}$ and $\tau_{max}$, respectively, for $i=1,2,3, \dots, N_{bins}$, where $N_{bins}$ is the total number of bins. The total number of sliding windows, $N_{bins}$, is defined by the ratio between the sampled range $\tau$ and the window spacing $\delta \tau$: $N_{bins} \simeq (\tau_{max}-\tau_{min})/ \delta \tau$ 
\footnote{To be more precise $N_{bins} = \rm{len}({\tau_i})$, where the centers are generated as $\tau_i = \tau_{min} + (i-1)\delta \tau$, being $\tau_i < \tau_{max}$.}.
Since $\delta \tau < \Delta \tau$, consecutive windows overlap, resulting in a sliding-window representation of the distribution. 
The resulting counts are treated as Poisson-distributed data with uncertainties $\sqrt{n_i}$, where $n_i$ is the number of persistence times (counts) that fall within the ith time window (bin) centered at $\tau_i$, namely the number of persistence times $\tau$ such that 
$\tau \in \left[\tau_i - \frac{\Delta\tau}{2},\ \tau_i + \frac{\Delta\tau}{2}\right)$.
The window centers $\tau_i$ are defined on a uniform grid, 
\begin{equation}
\tau_i = \tau_{min}+(i -1) \; \delta \tau, 
\end{equation}
with spacing $\delta \tau$, spanning the range $[\tau_{min}, \tau_{max})$ of the persistence times obtained in the simulation considered. 

Using the above procedure effectively in the simulation depicted in Fig.~\ref{fig:timeseries_b1}, we select the threshold $\tau>50$, a sliding-window width $\Delta \tau = 100$ and a window step $\delta \tau = 20$ (centers of the sliding windows are then defined), all expressed in dimensionless time units.
%
We then identify significant local maxima in $P(\tau)$ using a peak-finding algorithm and use their positions to initialize a maximum-likelihood fit based on a sum of Gaussian components.
The fit is performed by minimizing the Poisson negative log-likelihood under bounded parameters (positive amplitudes, constrained peak centers, and reasonable widths). 
As a result, we obtain the binned counts computed for the simulation in Fig.~\ref{fig:timeseries_b1}, along with the corresponding best-fit multi-Gaussian model displayed in Fig.~\ref{fig:Ptau_gaussians}, where the time center $\tau_n$ for each $n$-th Gaussian component is also reported in the legend.
%
\begin{figure*}[t]
    \centering
    \includegraphics[width=\textwidth]{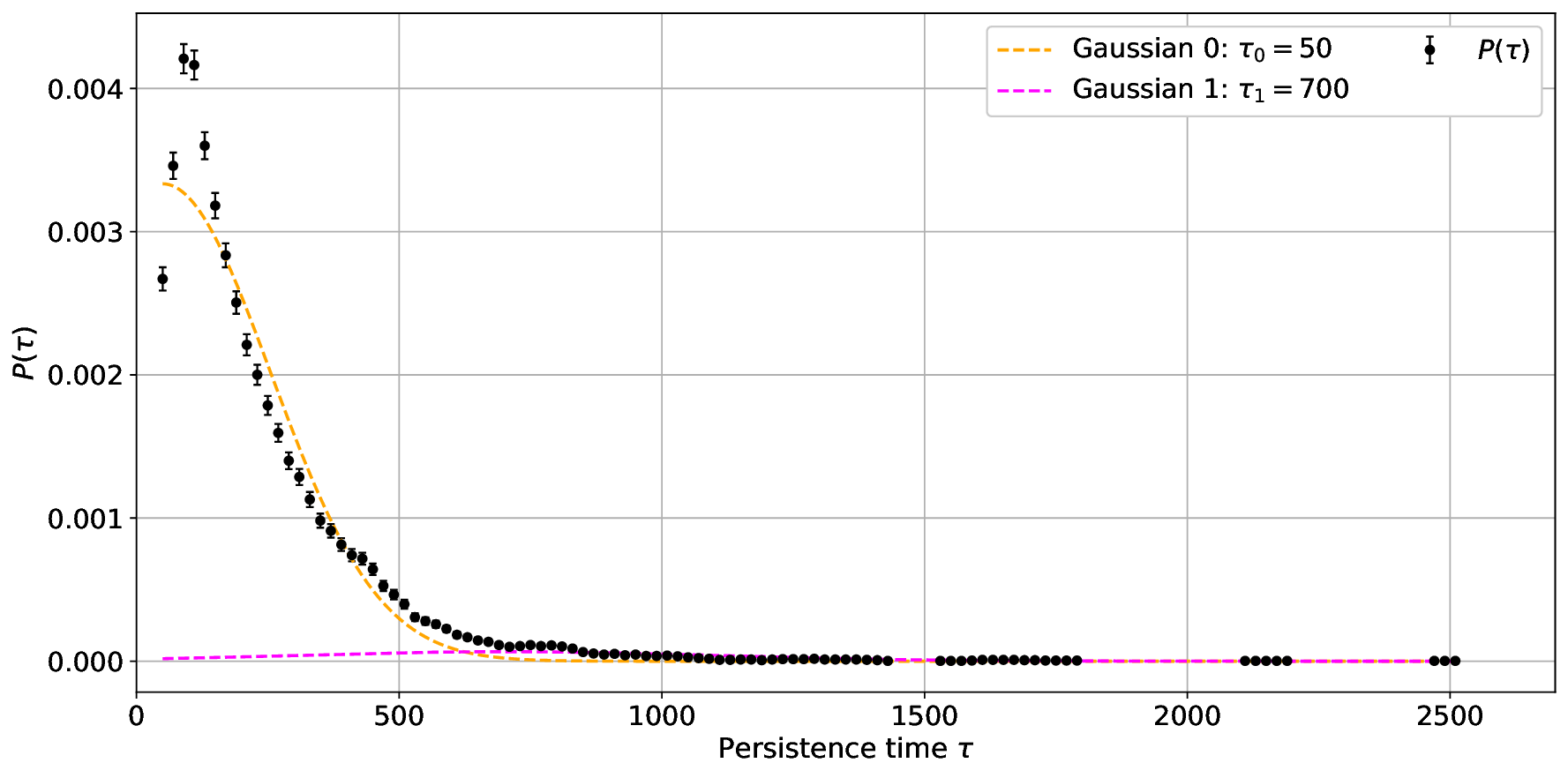}
    \caption{When the modulation of $\mu$ is removed, thereby eliminating the effect of kinetic-helicity modulation on the magnetic induction, stochastic resonance does not arise in our bistable model. Accordingly, the probability density function $P(\tau)$ no longer exhibits the multi-peak structure observed when this modulation is included. Basically, a single Gaussian is sufficient to fit the $P(\tau)$ function.}
    \label{fig:Ptau_gaussians_mu_constant}
\end{figure*}
It is not difficult to note that the local maxima are all almost very close to the multiple of the fundamental time $T_{\omega} = 500$ in dimensionless units, which is the period of the $\mu$-parameter function, that is,
\begin{equation}
\tau_n = n \; T_{\omega},
\label{eq:tau_n}
\end{equation}
with $n$ integer number (specifically, in this test case $n=1,2,3,\dots, 11$).
Equation (\ref{eq:tau_n}) holds for each local Gaussian maxima except for the maximum at time $\tau_0 \sim 100$.

For comparison, we repeat the same analysis previously made for the simulation in Fig.~\ref{fig:timeseries_b1}, now for the $\mu$-constant case, namely for the corresponding simulation obtained with $\mu = 1$, keeping unchanged all other parameters. We obtain a markedly different shape of $P(\tau)$ (see Fig.~\ref{fig:Ptau_gaussians_mu_constant}). In particular, for constant $\mu$, the distribution no longer exhibits the local Gaussian components previously observed, showing no statistically significant local maxima except for a single peak around $\tau_0$. This peak is also present in the corresponding $\mu$-modulated case. 
Since it persists when the periodic modulation is removed, and since in the constant-$\mu$ simulation the only remaining time scales are those generated by the nonlinear MHD shell dynamics, we interpret $\tau_0$ as an intrinsic time scale of the turbulent dynamics. This interpretation is also supported by the fact that the position of this peak does not scale with the imposed modulation period, as we tested by running other simulations with different $T_{\omega}$ values. For this reason, the peak at $\tau_0$ is excluded from the subsequent multi-peak analysis.

Therefore, the distribution $P(\tau)$ for the system with constant $\mu$ does not show statistically significant multimodality under the same analysis, suggesting the absence of distinct characteristic persistence-time scales, except $\tau_0$. This comparison clearly indicates that the characteristic timescales $\tau_n$ with $n=1,2,3 \dots$ identified when $\mu$ is time-modulated arise as a consequence of such modulation.

Subsequently, we define the peak height for each Gaussian component in the normalized distribution as
\begin{equation}
s_n \equiv P(\tau_n)=\frac{A_n}{N\,\Delta\tau},
\end{equation}
where $A_n$ is the fitted amplitude in counts, $N$ is the total number of events, and $\Delta\tau$ is the effective bin width. We assign an uncertainty using a Poisson-like estimate on the fitted peak counts,
\begin{equation}
\sigma_{s_n}=\frac{\sqrt{A_n}}{N\,\Delta\tau},
\end{equation}

\begin{figure}[t]
  \centering
    \includegraphics[width=\columnwidth]{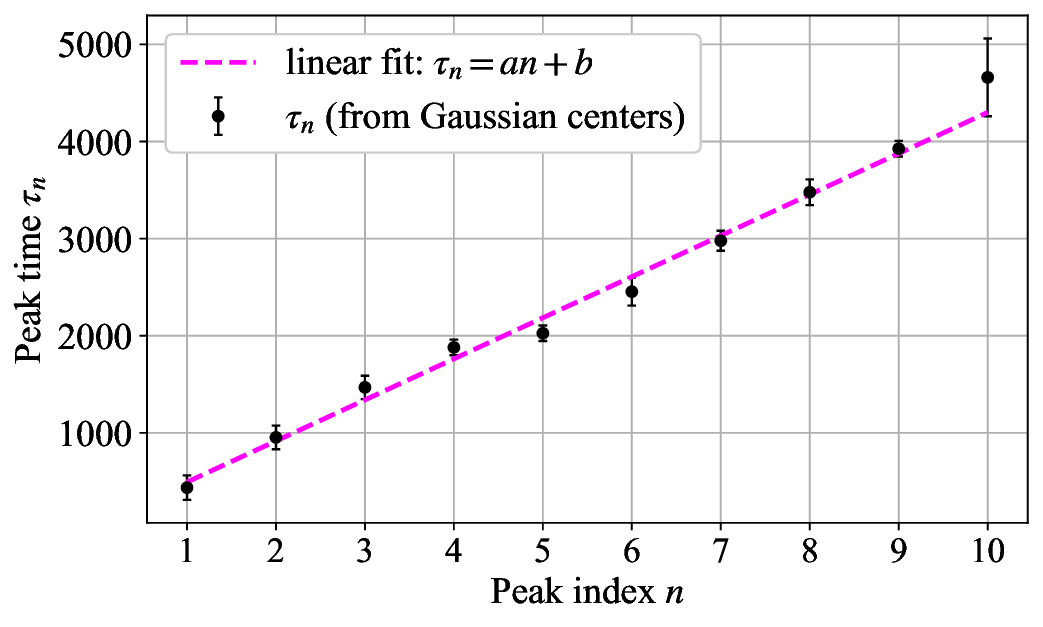}
  \caption{Peak times $\tau_n$ of the Gaussian centers versus index $n$ with linear fit $\tau_n = a n + b$, being $a = 423 \pm 13$ and $b = 71 \pm 80$. The error bars indicate the fitted Gaussian widths $\sigma_n$, used here as an effective measure of the temporal extent of each peak.}
  \label{fig:tau_n}
\end{figure}

We note that the first fitted peak at $\tau = \tau_0$ always appears in the multi-Gaussian fit inferred in different simulations obtained by varying values of model parameters, like, for instance, dissipative coefficients and/or the periodicity $T_{\omega}$ of the $\mu(t)$ function (this is not shown). Since the first peak appears almost at the same time $\tau_0 \simeq 100$ in each simulation independently of the period $T_{\omega}$, it is very likely not related to the modulation of the parameter $\mu(t)$. 
Therefore, after excluding the first peak (located around $\tau \simeq 100$) and the fourth Gaussian component, we renumber the remaining maxima with a consecutive index $n=1,2,3,\dots$. We then fit the dependence of $\tau_n$ on $n$ with a linear model,
\begin{equation}
\tau_n = a\,n + b ,
\end{equation}
where $a$ and $b$ are estimated using a weighted least-squares procedure. As an effective uncertainty of each peak location, we adopt the fitted Gaussian width $\sigma_n$, which provides a natural measure of the temporal extent of the corresponding maximum. The best-fit line is shown in Fig.~\ref{fig:tau_n} as a magenta dashed curve, together with the measured values of $\tau_n$ and their associated uncertainties.

These features, in particular, the emergence of multiple preferred persistence times in the $P(\tau)$ structure, associated with the $\mu$ modulation, are consistent with the phenomenology of stochastic-resonance \citep{benzi1981mechanism, gammaitoni1998stochastic}. Similar signatures have also been reported in the analysis of geomagnetic reversal records \citep{consolini2003stochastic}, suggesting that stochastic resonance may play a role in the dynamics of geomagnetic polarity reversals.

\section{Discussion and Conclusions}  

The results presented in this work show that a thermally driven low-dimensional geodynamo model can develop clear modulation-controlled structures in the persistence-time statistics of magnetic polarity reversals. In particular, when the parameter $\mu(t)$, which accounts for the effect of kinetic helicity on the large-scale magnetic induction, i.e., the $\alpha$ effect, is slowly modulated in time, the probability density function $P(\tau)$ of magnetic residence times displays a sequence of well-defined local maxima. By contrast, when modulation is removed and $\mu$ is kept constant, the multimodal structure disappears and the persistence-time distribution is reduced to a single-peak form. This comparison indicates that the characteristic timescales emerging in the modulated case are not a generic property of the turbulent shell dynamics alone but arise as a direct consequence of the slow periodic modulation imposed on the system.

A physical interpretation of these results is provided by the effective bistable structure of the large-scale magnetic dynamics. The reduced dynamics of the large-scale field $b_1(t)$ can be represented in terms of a double-well potential, with the two minima corresponding to the two preferred polarity states. In this framework, the time dependence of $\mu(t)$ acts as a slow modulation of the effective barrier that separates the two metastable states, while turbulent fluctuations generated by the nonlinear interactions between neighboring shells provide the perturbations that can trigger transitions between them. The coexistence of bistability, coherent modulation, and fluctuations naturally places the model within the class of systems that can exhibit stochastic-resonance-like behavior.

The persistence-time statistics support this interpretation. In the modulated regime, the local maxima of $P(\tau)$ are approximately aligned with the integer multiples of the modulation timescale $T_{\omega}$, and the fitted peak centers follow an approximately linear dependence on their index. This shows that the reversal dynamics does not remain characterized by a single broad waiting-time distribution, but instead becomes organized around preferred persistence-time scales selected by the external modulation. The disappearance of this structure in the constant-$\mu$ case further strengthens the conclusion that modulation controls the temporal organization of the reversals.

In the simplest periodically rocked double-well models of stochastic resonance, residence-time maxima are often discussed in relation to odd multiples of half the forcing period. In the present model, however, the modulation acts through the parameter $\mu(t)$, which controls the effective stability of the large-scale magnetic states by periodically lowering and raising the barrier between them. Within this setup, the most favorable conditions for polarity transitions are associated with the phases of minimum barrier height, which recur once per modulation cycle. This provides a natural explanation for why the dominant persistence-time scales identified in our simulations are approximately organized around integer multiples of $T_{\omega}$. 
This behavior is consistent with the multiplicative modulation mechanism discussed by Lorito et al. \citep{Lorito_et_al_2005AN}, in which periodic forcing acts symmetrically on the bistable structure by modulating the effective dynamo efficiency, rather than by asymmetrically tilting the two wells. In that case, the preferred residence times are expected around integer multiples of $T_{\omega}$, instead of odd multiples of $T_{\omega}/2$ typical of additive forcing. 

It is important to stress that modulation-controlled statistics do not imply deterministic reversal triggering. In our simulations, the $\mu(t)$ modulation does not prescribe the occurrence time of individual reversals. Instead, it periodically changes the stability of the two polarity states by lowering and raising the barrier between them. Reversals remain fluctuation-assisted events: they occur only when turbulent fluctuations are sufficiently strong to drive the large-scale magnetic mode across the unstable region near $b_1=0$. The modulation therefore biases the transition probability in time, producing preferred residence-time intervals, but individual reversals remain irregular events triggered by turbulent fluctuations.

From a geodynamo perspective, these results suggest that slow modulations of dynamo efficiency may leave a clear statistical imprint on reversal sequences, even in the presence of strong turbulent fluctuations. In this sense, the model supports the idea that long-time-scale modulations, possibly associated with changes in kinetic helicity due to CMB condition variations or other slow processes affecting the dynamics of the Earth's core, may affect the statistics of polarity persistence without directly imposing periodic or deterministic reversals. Here, by deterministic reversals we mean reversals whose occurrence time is directly prescribed by the external modulation. In our model, the reversals remain fluctuation-assisted events, but their timing is no longer statistically structureless: it is shaped by the interaction between bistability, turbulence, and slow coherent modulation. 

In this context, previously reported orbital-scale signatures in paleomagnetic records acquire a possible dynamical interpretation. Yamazaki and Oda reported a periodicity of $\sim 100$-kyr in inclination and intensity, suggesting that the geomagnetic field may be modulated at orbital timescales \citep{Yamazaki_Oda_Science_2002, Yamazaki_PEPI_2002}.  Channell et al. also found evidence for orbital-scale variability in paleointensity records \citep{Channell_et_al_Nature_1998}. 

Our results do not demonstrate that orbital forcing is the unique or direct cause of reversals; rather, they show that if long-period modulations are transmitted to the dynamo through slowly varying control parameters, the reversal sequence can acquire preferred persistence-time scales in a way that is fully compatible with a bistable stochastic-resonance-like scenario. 
In terms of our description, orbital variations could influence the dynamics of outer-core magnetoconvection and thus alter the kinetic helicity, which in turn can increase or decrease the probability of geomagnetic reversals. 
In general, possible physical realizations of such slow modulations include changes in the CMB heat flux \citep{Glatzmaier_et_al_1999, Olson_Christensen_GeoJI_2002, Nakagawa_Tackley_GGG_2005}, thermochemical heterogeneity in the lowermost mantle \citep{Garnero_McNamara_Sci_2008, Nakagawa_Tackley_GeoRL_2011, Olson_GGG_2016}, or other processes capable of slowly modifying the vigor of core convection \citep{Olson_et_al_2010, Driscoll_Olson_2011, Biggin_et_al_2012, Olson_et_al_2014E&PSL}.

In conclusion, the findings we discussed in this paper suggest that the geodynamo may respond to long-timescale modulation by organizing the statistics of fluctuation-assisted transitions between metastable polarity states. In this sense, the present study provides a natural bridge between reduced bistable descriptions of polarity reversals, stochastic-resonance interpretations of paleomagnetic reversal statistics, and geodynamical scenarios in which slow variations in dynamo efficiency -- possibly linked to variations in CMB heat flux or other orbital-scale modulation -- affect the stability of the magnetic dipole state.

\begin{acknowledgments}
This work is supported by the MELODY research project, funded by the Italian Ministry of Universities and Research (MUR) under the PNRR Young Researchers program 2022 (MELODY SoE project, grant agreement No SOE\_0000119, CUP E53C22002450006).
\end{acknowledgments}

\bibliography{biblio}

\end{document}